\documentclass[
    aip,
    reprint,
    amsmath,
    amssymb,
    superscriptaddress,
    floatfix
]{revtex4-2}

\usepackage{graphicx}
\usepackage{dcolumn}
\usepackage{bm}
\usepackage{hyperref,physics}
\usepackage{xcolor}

\begin{document}

\title{Quantum magnonics with engineered dissipative and coherent interactions}

\author{Debsuvra Mukhopadhyay}
\thanks{Equal contributor.}
\email{debsuvram@usf.edu}
\affiliation{Department of Physics, University of South Florida, Tampa, Florida 33620, USA.}

\author{Jayakrishnan M. P. Nair}
\thanks{Equal contributor.}
\email{jaynair@udel.edu}
\affiliation{Department of Physics and Astronomy,
University of Delaware, Newark, Delaware 19716, USA.}

\author{Girish S. Agarwal}
\email{girish.agarwal@tamu.edu}
\affiliation{Institute for Quantum Science and Engineering,
Texas A\&M University, College Station, Texas 77843, USA.}
\affiliation{Department of Biological and Agricultural Engineering, Texas A\&M University, College Station, Texas 77843, USA.}

\date{\today}

\begin{abstract}
Cavity magnonics is moving beyond the conventional paradigm of coherent spin--photon hybridization toward the controlled engineering of dissipative and coherent interactions, Floquet dynamics, and quantum fluctuations. This Perspective examines how these ingredients reshape collective dynamics and open new avenues for quantum state engineering, transport, and sensing. We discuss the role of dissipative coupling and anti-$\mathcal{PT}$ symmetry in enabling long-lived collective modes and enhanced spectral response, and highlight Floquet engineering as a means of engineering nonreciprocal transport and phase-controlled interactions. We then review recent progress in quantum magnonics, ranging from squeezed and entangled magnon states, including the squeezing of thermal magnons, to prospective schemes for preparing magnonic Schr\"odinger-cat states. We also discuss how parametric driving and active gain strongly modify hybrid susceptibilities and amplify magnon--photon response. These advances expand the scope of cavity magnonics beyond conventional polariton physics, establishing it as a versatile platform for sensing, transduction, spin transport, and hybrid quantum technologies.
\end{abstract}

\maketitle

\section{Introduction}
\label{sec:Intro}

Cavity magnonics has developed into a broad platform for manipulating collective spin excitations via their interaction with electromagnetic fields. Early demonstrations of strong magnon--photon coupling \cite{Huebl2013,Tabuchi2014,Zhang2014,Goryachev2014} established the coherent hybridization of magnons and microwave photons into cavity--magnon polaritons. Since then, the field has expanded well beyond normal-mode splitting, with increased attention to engineered dissipation, nonlinear response, quantum fluctuations, and driven nonequilibrium phenomena
\cite{RameshtiPhysRep2022,YuanPhysRep2022}.

Much of this flexibility originates from low-loss ferrimagnets such as yttrium iron garnet (YIG), which have emerged as versatile candidates for cavity-magnonic experiments. Their high spin density enables strong collective coupling, while weak magnetic damping supports long coherence times. The same material also affords intrinsic nonlinear and multimode interactions: magnetocrystalline anisotropy produces magnon Kerr nonlinearities, while magnetostriction couples magnons to mechanical motion \cite{Zhang2016,LiPRL2018}, and magneto-optical interactions connect microwave and optical degrees of freedom \cite{RameshtiPhysRep2022, YuanPhysRep2022}. Together, these properties
allow cavity-magnonic systems to access regimes that are difficult to realize with purely linear spin--photon coupling.

A particularly promising development has been the extension of cavity magnonics into nonlinear and quantum regimes. Coherent coupling to superconducting qubits has linked collective spin excitations directly to circuit QED \cite{Tabuchi2015,LachanceQuirion2019}, while engineered reservoirs can mediate correlated decay and produce level attraction, dark collective modes, and non-Hermitian dynamics \cite{Harder2018,Nair2021Sensing}. Such dissipative interactions can also entail microwave directionality \cite{WangPRL2019Nonreciprocity} and enhance hybrid transduction
\cite{Mukhopadhyay2022Conversion}. Intrinsic magnon nonlinearities lead to multistability in multimode systems \cite{Wang2018Bistability,Shen2021Multistability} and have been shown to reshape cavity-mediated spin transport \cite{Nair2020SpinCurrents}, building on the demonstration of photon-mediated spin-current transfer between spatially separated magnetic samples \cite{Bai2017}. Parametric driving offers an additional control knob for modifying the hybrid susceptibility and amplifying cavity-mediated response \cite{MukhopadhyayPRB2022Amplification}.

These developments have advanced cavity magnonics increasingly closer to the quantum regime. Recent work has explored the preparation of Fock, squeezed, entangled, and cat-like magnon states, together with their possible use in quantum information, sensing, and studies of macroscopic quantum behavior \cite{lu2026quantum}. Strong-dispersive qubit--magnon coupling has enabled magnon-number-resolved spectroscopy \cite{LachanceQuirion2017}, while the deterministic preparation and coherent control of single-magnon states have now been demonstrated \cite{Xu2023}. Several theoretical proposals have also highlighted how squeezed and entangled magnon states could be generated using magnomechanical interactions \cite{LiPRL2018,Li2019Squeezing}, intrinsic magnon nonlinearities \cite{Zhang2019KerrEntanglement}, or the transfer of squeezing from microwave fields to magnons \cite{Nair2020Deterministic}.

In this Perspective, we focus on four directions that illustrate how cavity magnonics is advancing beyond conventional coherent hybridization. In Sec. \ref{sec:dissipative}, we discuss dissipative coupling and anti-$\mathcal{PT}$ symmetry, emphasizing long-lived collective modes and their implications for nonlinearity, sensing, and transduction. Sec. \ref{sec:Floquet} examines Floquet engineering for nonreciprocity, asymmetric mode coupling, and directional transport. In Sec. \ref{sec:quantum}, we review recent advances in quantum magnonics, including squeezed and entangled states, and outline prospective routes toward preparing Schr\"odinger-cat states. Finally, in Sec. \ref{sec:parametric}, we consider parametric driving and gain, emphasizing their roles in amplification, critical response, quantum-fluctuation engineering, as well as quantum sensing. Collectively, these developments illustrate how reservoir engineering, periodic driving, nonlinearity, and quantum control are extending the capabilities of cavity magnonics across classical and quantum regimes, and motivate the opportunities and challenges discussed in Sec. \ref{sec:outlook}.

\begin{figure}[t]
\centering
\includegraphics[width=0.95\columnwidth]{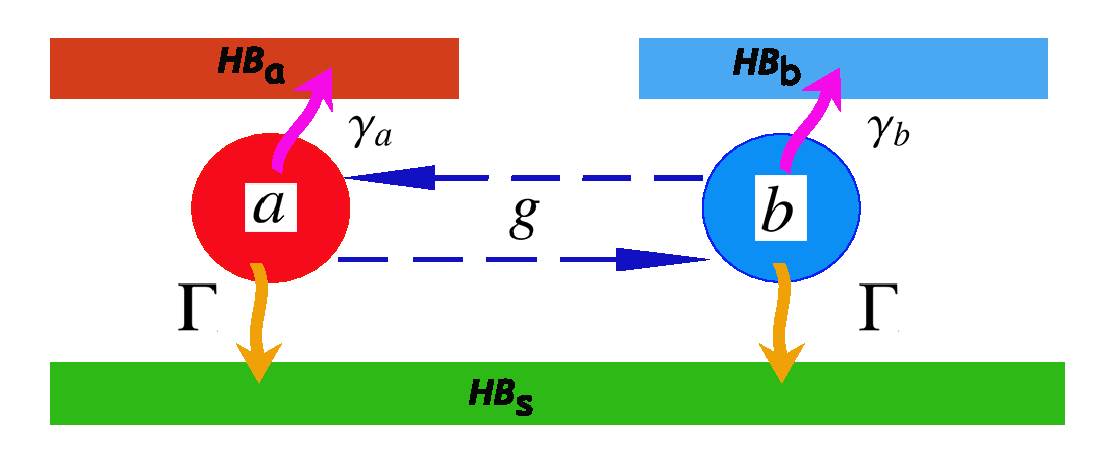}
\caption{Schematic of a generic dissipatively coupled two-mode system. Each mode interacts with its local environment while simultaneously coupling to a common reservoir, which mediates an effective dissipative interaction between the modes through correlated decay. This generic configuration forms the basis for a variety of non-Hermitian phenomena, including level attraction, anti-$\mathcal{PT}$ symmetry, and EP physics. Adapted from Ref.~\cite{Nair2021Sensing}.}
\label{fig:dissipative_setup}
\end{figure}

\section{Cavity magnonics and enhanced quantum sensing via Anti-$\mathcal{PT}$ symmetry induced long lived mode}
\label{sec:dissipative}

Cavity magnonics is established as a platform for investigating coherent interactions between magnons and microwave photons, including strong coupling and the formation of hybrid cavity--magnon polaritons. An alternate coupling mechanism involves the use of an intermediary electromagnetic environment to mediate correlated decay between magnetic modes, turning dissipation itself into an interaction resource \cite{Harder2018}. Such reservoir-mediated dissipative coupling provides access to level attraction, exceptional points (EPs), and non-Hermitian collective dynamics that have now become equally profound themes in cavity magnonics \cite{RameshtiPhysRep2022}.  A typical realization is illustrated in Fig.~\ref{fig:dissipative_setup}, where spatially separated magnon modes couple to a common waveguide that mediates their effective dissipative interaction.

For two magnon modes coupled to a common cavity, the familiar strong coherent coupling regime permits a simplified picture in terms of bright and dark collective superpositions: the bright mode couples to the cavity and forms polaritons, whereas the orthogonal dark mode remains decoupled \cite{NairPRB2022Levels}. A qualitatively different outcome emerges in the bad-cavity limit wherein the cavity relaxes rapidly compared with the magnons. Adiabatic elimination of the cavity then gives the effective inter-magnon self-energy
\begin{align}
\Sigma_{12}
&=
\frac{g_1g_2}{\Delta_c+i\kappa/2}
\nonumber\\
&=
\frac{g_1g_2\Delta_c}
{\Delta_c^2+(\kappa/2)^2}
-
i\frac{g_1g_2(\kappa/2)}
{\Delta_c^2+(\kappa/2)^2},
\label{eq:cavity_mediated_self_energy}
\end{align}
where $\Delta_c$ and $\kappa$ are the cavity detuning and linewidth, and $g_{1,2}$ are the coupling strengths. The real part is equivalent to a coherent cavity-mediated exchange, whereas the imaginary part represents correlated dissipation through the common electromagnetic reservoir. Near cavity resonance, the latter can dominate, allowing the cavity to act primarily as a dissipative mediator \cite{NairPRB2022Levels}.

Dissipative coupling provides a natural setting for anti-$\mathcal{PT}$ symmetry. While a conventional $\mathcal{PT}$-symmetric Hamiltonian commutes with the combined parity--time operation \cite{Bender1998}, an anti-$\mathcal{PT}$ Hamiltonian anticommutes with it after removal of a constant loss term \cite{Ge2013AntiPT,Yang2020AntiPT}. For identical magnon losses and predominantly dissipative coupling, the effective two-mode Hamiltonian takes the form
\begin{equation}
\frac{H_{\rm eff}}{\hbar}
=
\begin{pmatrix}
\delta-i\gamma & -i\Gamma\\
-i\Gamma & -\delta-i\gamma
\end{pmatrix},
\qquad
\lambda_\pm
=
-i\gamma
\pm
\sqrt{\delta^2-\Gamma^2},
\label{eq:anti_pt_spectrum}
\end{equation}
where $\delta$ is the relative magnon detuning, $\gamma$ the common overall decay rate, and $\Gamma$ the dissipative coupling strength. This spectral structure has been theoretically explored in dissipatively coupled cavity-magnonic systems \cite{Nair2021Sensing}.

\begin{figure}[t]
\centering
\includegraphics[width=0.95\columnwidth]{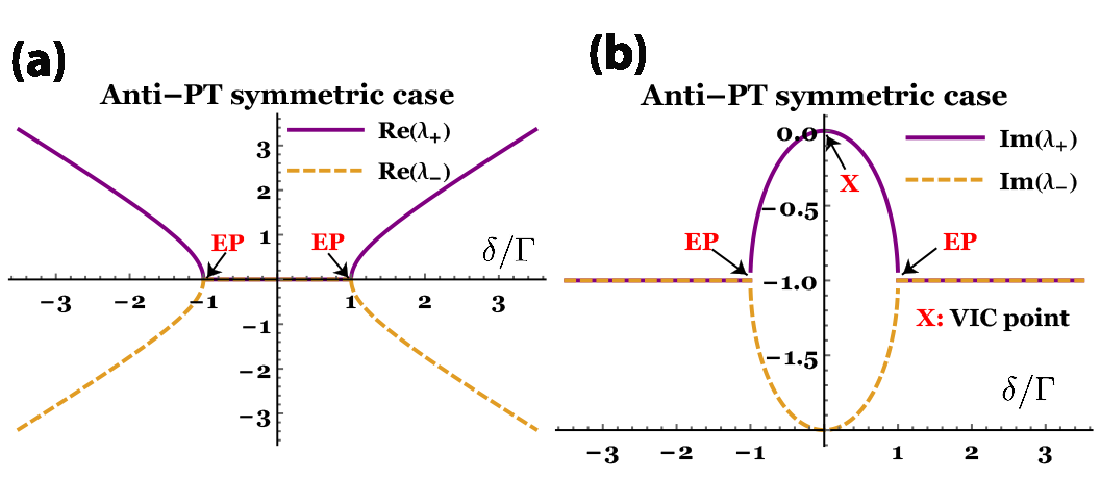}
\caption{Characteristic eigenspectrum of a dissipatively coupled anti-$\mathcal{PT}$-symmetric system. (a) Real and (b) imaginary parts of the eigenvalues as functions of the relative detuning $\delta$. The EPs at $|\delta|=\Gamma$ separate the spectrally split regime from the regime of frequency coalescence and linewidth bifurcation into superradiant and subradiant branches. Adapted from Ref.~\cite{Nair2021Sensing}.}
\label{fig:antiPT}
\end{figure}

As illustrated in Fig.~\ref{fig:antiPT}, for $|\delta|>\Gamma$ the two resonances remain frequency split while having the same linewidth. At $|\delta|=\Gamma$, the eigenvalues and eigenvectors coalesce at EPs. For $|\delta|<\Gamma$, the frequencies merge while the linewidths separate into superradiant and subradiant branches, with the resonant ($\delta=0$) linewidths becoming $\gamma+\Gamma$ and $\gamma-\Gamma$, respectively. The resulting subradiant excitation can therefore survive substantially longer than the bare magnon mode, concentrating the response into a narrow collective resonance. This long-lived mode has been exploited as a resource for several nonlinear and transduction phenomena discussed below \cite{Nair2021Ultralow,Mukhopadhyay2022Conversion}.

The subradiant linewidth can be particularly consequential in the presence of magnetic Kerr nonlinearities. It can be shown that the onset of bistability scales approximately as
\begin{equation}
|\Omega_{\rm th}|^2
\propto
\frac{\gamma_{\rm eff}^{\,3}}{K},
\qquad
\gamma_{\rm eff}\simeq\gamma-\Gamma,
\label{eq:dissipative_bistability_threshold}
\end{equation}
where $K$ denotes the Kerr strength. Consequently, even a modest reduction of the collective linewidth can greatly lower the required drive power. The resulting ultralow-threshold bistability has been developed theoretically in Refs.~\cite{Nair2021Ultralow} and subsequently explored experimentally \cite{Pan2022Bistability,Pan2022Bistability}. The same enhanced susceptibility can be exploited for sensing weak magnetic perturbations. Employing a long-lived mode, even a small Kerr shift can produce a pronounced change in the steady-state magnon population and the associated spin-current response \cite{Nair2021Sensing}. 

The connection to parameter-estimation precision can be sharpened by using quantum Fisher information \cite{Wang2022QFI}. In the effective two-magnon anti-$\mathcal{PT}$ description, let $\alpha_0(\varepsilon)$ and $\beta_0(\varepsilon)$ denote the complex steady-state magnon amplitudes in response to a microwave drive, where $\varepsilon$ represents a weak magnetic perturbation, such as a detuning mismatch or an additional coherent coupling. For the long-time coherent state, the quantum Fisher information simplifies exactly as
\begin{equation}
\mathcal{F}_Q(\varepsilon)
=
4\left|
\frac{\partial\alpha_0}{\partial\varepsilon}
\right|^2
+
4\left|
\frac{\partial\beta_0}{\partial\varepsilon}
\right|^2.
\label{eq:antiPT_QFI}
\end{equation}
Thus, the large derivative of the magnonic response near the long-lived anti-$\mathcal{PT}$ resonance translates directly into enhanced parameter sensitivity. In the near-singular regime considered in Ref.~\cite{Wang2022QFI}, the response derivatives scale as $\varepsilon^{-2}$, yielding $\mathcal{F}_Q\propto\varepsilon^{-4}$ and $\delta\varepsilon_{\rm CRB}\propto\varepsilon^2$.

Reservoir engineering can likewise benefit hybrid quantum interfaces. A dissipatively coupled microwave cavity and ferromagnetic mode, combined with magneto-optical coupling to an optical cavity, provides a route to enhanced microwave--optical conversion \cite{Mukhopadhyay2022Conversion}. Operating near the anti-$\mathcal{PT}$ regime increases the effective magnon susceptibility and strengthens the coherence linking the microwave, magnonic, and optical degrees of freedom, thereby improving transduction\cite{Mukhopadhyay2022Conversion}. Dissipative coupling can also lead to directional behavior. In multiport geometries, propagation-phase-sensitive interference between coherent and dissipative scattering pathways can produce different forward and backward transmission amplitudes, including unidirectional invisibility \cite{WangPRL2019Nonreciprocity}.

\section{Floquet engineering}
\label{sec:Floquet}

\begin{figure}[t]
\centering
\includegraphics[width=1\columnwidth]{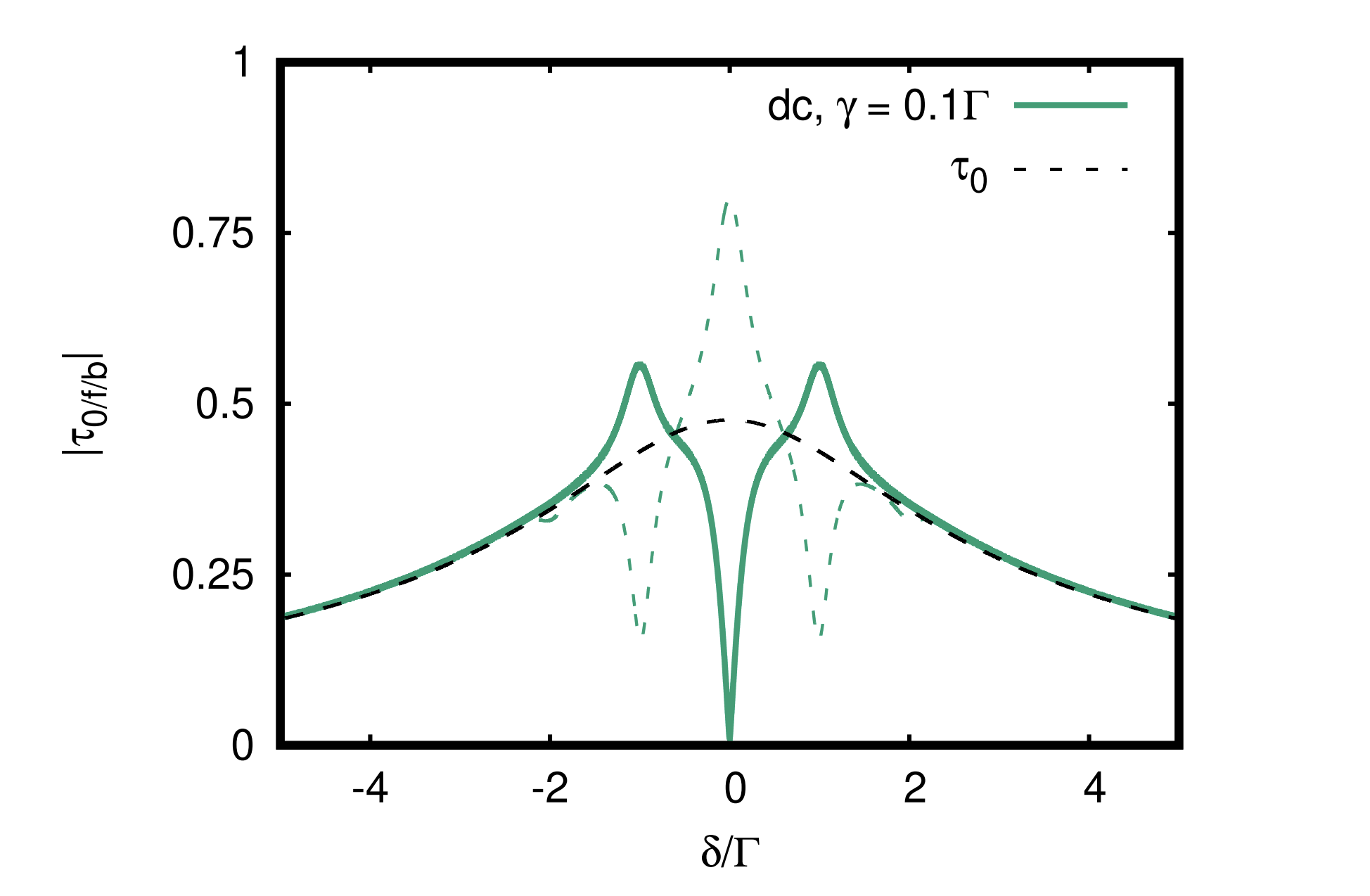}
\caption{Floquet-induced nonreciprocal transmission in dissipatively coupled resonators. The forward (solid green) and backward (dashed green) transmission amplitudes are shown as functions of the normalized detuning $\delta/\Gamma$, where $\delta=\omega_0-\omega_l$ and $\Gamma$ is the dissipative coupling rate. The black dashed curve shows the transmission $\tau_0$ without modulation. The parameters are $\Omega=\Gamma$, $\beta=\Gamma$, $\theta=\pi/2$, and $\gamma=0.1\Gamma$, wwhere $\Omega$ and $\beta$ are the modulation frequency and amplitude, respectively, and $\theta$ is the relative phase between the frequency modulations of the two resonators. Adapted from Ref.~\cite{Biehs2023}.}
\label{fig:floquet}
\end{figure}

Periodic driving provides a versatile route to nonreciprocal transport by harnessing interference among Floquet sidebands. In two dissipatively coupled resonators, frequency modulation with a relative phase generates a synthetic magnetic field in the Floquet sideband space, thereby breaking reciprocity between forward and backward propagation \cite{Biehs2023}. As shown in Fig.~\ref{fig:floquet}, the resulting transmission exhibits pronounced directional contrast around the central resonance and the Floquet sidebands. In particular, transmission in one direction can be enhanced while that in the opposite direction is suppressed, with complete suppression in one direction possible at zero detuning. Here, $\delta=\omega_0-\omega_l$ is the detuning between the common resonator frequency $\omega_0$ and the driving-field frequency $\omega_l$, while $\Gamma$ denotes the dissipative coupling rate. This nonreciprocal behavior is strongly enhanced by the long-lived mode supported by dissipative coupling, which also substantially narrows the directional transmission features.

More broadly, Floquet engineering provides a powerful technique for controlling interactions in magnonic systems. Periodic modulation has recently been used to coherently couple multiple magnon modes in frequency space, realizing a reconfigurable synthetic magnonic lattice and spectral Bloch oscillations \cite{jtpv-d5k6}. In addition, dual-tone Floquet modulation of a cavity--magnon system can create interference between distinct Floquet pathways, allowing the relative drive phase to selectively enhance one Floquet-mediated coupling channel while suppressing the opposite one and thereby producing phase-programmable asymmetric mode coupling \cite{pishehvar2026asymmetric}. 

Yet another avenue is afforded by magnonic nonlinearity: Kerr-induced direction-dependent switching can place counterpropagating signals on different bistable branches \cite{Miao2024}. Further, chiral waveguide QED can extend this directionality to photon statistics and quantum fluctuations as well \cite{Miao2025}. These results suggest that Floquet-engineered phases, reservoir-mediated interference, and magnetic nonlinearities could be combined to engineer power-dependent and, ultimately, quantum directional transport in cavity-magnonic platforms.

Finally, these developments highlight dissipation as an engineered resource for various hybrid magnon-photon interfaces. By redistributing decay among collective magnon modes, a shared reservoir can generate long-lived excitations that enhance nonlinear response, directionality, perturbation sensitivity, and transduction. Dissipatively coupled cavity magnonics, therefore, provides a complementary route to controlling hybrid spin–photon systems, especially in regimes where coherent coupling alone may be insufficient.

\section{Quantum Magnonics}
\label{sec:quantum}

\subsection{Entanglement and Squeezing in Magnonic systems}

The collective quantum behavior of magnons makes the preparation of nonclassical states particularly compelling, as these quasiparticles describe the coherent dynamics of a macroscopic ensemble of microscopic spins. Recent experiments have established the basic ingredients for quantum magnonics by demonstrating quantum control at the single-magnon level, including the resolution of individual magnon-number states \cite{LachanceQuirion2017} and the deterministic preparation and coherent manipulation of single-magnon quantum states\cite{Xu2023}. These advances have stimulated considerable theoretical interest in generating nontrivial nonclassical states, such as squeezed and entangled magnons, by exploiting the wide variety of interactions available in hybrid cavity-magnonic systems.

One of the most widely used platforms is cavity magnomechanics, where coherent magnon--photon and magnon--phonon interactions are harnessed to distribute quantum correlations among electromagnetic, magnetic, and mechanical degrees of freedom. In a YIG sphere, magnons couple to microwave cavity photons through the magnetic-dipole interaction, while magnetostriction couples the magnon to mechanical vibrations through the nonlinear interaction
\begin{equation}
H_{\rm mp}
=
\hbar g_{mb}
m^\dagger m
(b+b^\dagger),
\end{equation}
where $b$ denotes the mechanical mode. Under strong coherent driving, this interaction can be linearized, generating effective beam-splitter and two-mode squeezing interactions between the hybrid modes. Consequently, quantum correlations can be distributed among the different modes of the hybrid system, giving rise to steady-state bipartite and genuine tripartite entanglement \cite{LiPRL2018}. The same magnetostrictive interaction has also been proposed as a means of generating steady-state entanglement between two spatially separated magnon modes coupled to a common cavity \cite{LiZhu2019}.

An alternative mechanism relies on transferring quantum correlations from an externally prepared field to the magnonic system. When the cavity is driven by squeezed microwave vacuum, generated, for example, by a Josephson parametric amplifier, the coherent cavity--magnon interaction
\begin{equation}
H_{\rm int}
=
\hbar g
\left(
a^\dagger m
+
a m^\dagger
\right)
\end{equation}
transfers the reduced fluctuations of the squeezed cavity field to the magnon mode\cite{Li2019Squeezing}. In a cavity-magnomechanical system, a red-detuned magnon drive can activate an effective beam-splitter interaction between magnons and phonons, allowing the transferred squeezing to be shared with the mechanical mode. The same principle can be extended to two spatially separated YIG samples, with the cavity acting as a quantum bus that maps the squeezing of the incident microwave field onto two collective magnon modes. This leads to deterministic magnon--magnon entanglement without relying on intrinsic Kerr or magnetostrictive nonlinearities \cite{Nair2020Deterministic}. This protocol serves as an explicit example of reservoir engineering where coherent state transfer constitutes the primary resource for entanglement generation.

It is also possible to generate the squeezed correlations directly within the magnetic system. Recent experiments in YIG films have demonstrated single-mode thermal squeezing of magnetization fluctuations under degenerate parametric pumping, together with two-mode thermal squeezing produced by nondegenerate excitation of two phase-locked magnon modes \cite{Hioki2026}. In the latter case, parametric downconversion correlates magnons at distinct frequencies whose amplitudes are preferentially localized near opposite surfaces of the magnetic film. Although the observed fluctuations remain in the thermal rather than quantum regime, this result experimentally establishes direct phase-sensitive control of correlated magnon fluctuations and suggests a different route toward magnon--magnon entanglement in which the magnetic system itself supplies the squeezing interaction.

Yet another route exploits the magnetic system itself. The relevant Hamiltonian of this anharmonic term goes as
\begin{equation}
H_K
=
\hbar K
(m^\dagger m)^2,
\end{equation}
which is typically weak at the single-magnon level but can be substantially enhanced under strong coherent driving. Linearization about the driven steady state generates effective two-mode squeezing interactions between the cavity-coupled magnon modes, leading to steady-state entanglement between spatially separated magnetic systems \cite{Zhang2019KerrEntanglement}. The interplay between Kerr nonlinearity, coherent driving, and cavity-mediated coupling has also been predicted to produce bistable squeezed and entangled steady states \cite{Yang2021BistableQuantum}. These studies demonstrate that weak intrinsic magnetic nonlinearities can be transformed into effective quantum resources through nonequilibrium driving.

The internal structure of the magnetic material itself is also responsible for generating nonclassical correlations. In antiferromagnets, the exchange interaction naturally gives rise to non-number-conserving terms following the Holstein--Primakoff transformation,
\begin{equation}
H_{\rm AFM}
\sim
g_{ab}
(a b+a^\dagger b^\dagger),
\end{equation}
which results in two-mode squeezing. As a consequence, the antiferromagnetic ground state is intrinsically entangled. Coupling the antiferromagnet to a cavity has been predicted to further enhance this magnon--magnon entanglement through cavity-induced cooling toward the correlated ground state \cite{Yuan2020AFMEntanglement}. This mechanism illustrates that the cavity can not only mediate interactions between excitations but also amplify quantum correlations already present within the magnetic medium.

\subsection{Future perspective: Magnonic Schr\"odinger-cat states}
\label{sec:cat}

While most of the above approaches primarily target Gaussian nonclassical states, coupling magnons to a two-level quantum system \cite{Tabuchi2015,LachanceQuirion2017} introduces a qualitatively different resource. The strong anharmonicity of a superconducting qubit enables genuinely non-Gaussian dynamics and opens the possibility of preparing macroscopic superposition states of collective spin excitations. We next propose two prospective routes toward engineering magnon Schr\"odinger-cat states based on dispersive conditional-phase evolution and resonant Jaynes--Cummings dynamics. The first uses a dispersive qubit--magnon interaction, analogous to the conditional-phase protocols developed in atomic cavity QED \cite{Brune1992,Davidovich1996,Brune1996}. The second exploits the
number-dependent dynamics of the resonant Jaynes--Cummings model, following the phase-space analysis of Eiselt and Risken \cite{Eiselt1989,Eiselt1991}. In both approaches, the magnon mode is first
prepared in a coherent state $\ket\alpha$ by a microwave pulse and is then split into distinct components in phase space.

\subsubsection{Dispersive qubit--magnon interaction}

Consider a magnon mode of frequency \(\omega_m\) coupled to a superconducting qubit of frequency $\omega_q$,
\begin{equation}
\frac{H}{\hbar}
=
\omega_m m^\dagger m
+
\frac{\omega_q}{2}\sigma_z
+
g\left(
m^\dagger\sigma_-+m\sigma_+
\right).
\label{eq:qubit_magnon_JC}
\end{equation}
For a qubit--magnon detuning
\(\Delta=\omega_q-\omega_m\) satisfying
\begin{equation}
|\Delta|
\gg
g\sqrt{n+1}
\label{eq:magnon_dispersive_condition}
\end{equation}
over the relevant range of magnon occupations, direct excitation exchange is suppressed. To second order in \(g/\Delta\), after removing the renormalized bare frequencies, the relevant interaction is given by
\begin{equation}
\frac{H_{\rm int}}{\hbar}
=
\chi\,m^\dagger m\,\sigma_z.
\label{eq:conditional_magnon_phase}
\end{equation}
The magnon phase consequently evolves in opposite directions for the two qubit states.

This regime is particularly relevant to quantum magnonics because strong dispersive coupling between a superconducting qubit and a magnetostatic mode has already been demonstrated experimentally. Magnon-number states were resolved through their dispersive shifts of the qubit transition, establishing the number sensitivity required for conditional-phase control \cite{LachanceQuirion2017}.

Suppose that a coherent microwave pulse prepares the magnon mode in
\begin{equation}
|\alpha\rangle
=
e^{-|\alpha|^2/2}
\sum_{n=0}^{\infty}
\frac{\alpha^n}{\sqrt{n!}}|n\rangle,
\label{eq:magnon_coherent_state}
\end{equation}
while a Ramsey pulse prepares the qubit in \((|e\rangle+|g\rangle)/\sqrt{2}\). The joint initial state is given by
\begin{equation}
|\Psi(0)\rangle
=
\frac{1}{\sqrt{2}}
\left(
|e\rangle+|g\rangle
\right)|\alpha\rangle.
\label{eq:initial_qubit_magnon_state}
\end{equation}
Evolution under Eq.~\eqref{eq:conditional_magnon_phase} gives
\begin{equation}
|\Psi(t)\rangle
=
\frac{1}{\sqrt{2}}
\left[
|e\rangle
\left|\alpha e^{-i\chi t}\right\rangle
+
|g\rangle
\left|\alpha e^{i\chi t}\right\rangle
\right].
\label{eq:conditional_magnon_rotation}
\end{equation}
The qubit thus entangles with two coherent magnon states that rotate in opposite directions in phase space. For an interaction time
\begin{equation}
\chi t_{\rm cat}
=
\frac{\pi}{2},
\label{eq:dispersive_cat_time}
\end{equation}
the two components become \(|-i\alpha\rangle\) and \(|i\alpha\rangle\). A second Ramsey rotation maps their relative phase onto the qubit population, and a subsequent qubit measurement projects the magnon mode onto one of the parity-resolved cat states
\begin{equation}
|\mathcal{C}_{\pm}\rangle
=
\mathcal{N}_{\pm}
\left(
|i\alpha\rangle
\pm
|-i\alpha\rangle
\right),
\label{eq:magnonic_cat_states}
\end{equation}
with
\begin{equation}
\mathcal{N}_{\pm}
=
\left[
2\left(
1\pm e^{-2|\alpha|^2}
\right)
\right]^{-1/2}.
\label{eq:magnonic_cat_normalization}
\end{equation}

The feasibility of this protocol rests on a clear set of competing requirements. The interaction time $t_{\rm cat}=\pi/(2|\chi|)$ must be shorter than both the magnon coherence time and the relevant qubit dephasing time, while the dispersive condition in Eq.~\eqref{eq:magnon_dispersive_condition} must remain valid across the Poissonian number distribution of the initial coherent state. Increasing \(|\alpha|\) improves the distinguishability of the two components through the overlap
\(
|\langle i\alpha|-i\alpha\rangle|=e^{-2|\alpha|^2}
\),
but simultaneously increases their vulnerability to magnon loss and
dephasing.

\subsubsection{Resonant Jaynes--Cummings route}

A second route may exploit the resonant qubit--magnon exchange rather than dispersive phase accumulation. In the interaction picture, the dynamics are governed by the Jaynes--Cummings Hamiltonian
\begin{equation}
H_{\rm JC}
=
\hbar g
\left(
m^\dagger\sigma_-
+
m\sigma_+
\right).
\label{eq:magnon_JC_hamiltonian}
\end{equation}
For an initially excited qubit and a magnon mode prepared in a coherent state, the joint state is
\begin{equation}
|\Psi(0)\rangle
=
|e\rangle|\alpha\rangle
=
\sum_{n=0}^{\infty}
C_n|e,n\rangle,
\qquad
C_n
=
e^{-|\alpha|^2/2}
\frac{\alpha^n}{\sqrt{n!}},
\label{eq:initial_resonant_JC_state}
\end{equation}
with mean magnon occupation \(\bar n=|\alpha|^2\). Its exact evolution is
\begin{equation}
\begin{aligned}
|\Psi(t)\rangle
=
\sum_{n=0}^{\infty}C_n
\Big[
&
\cos\!\left(gt\sqrt{n+1}\right)|e,n\rangle
\\
&-
i\sin\!\left(gt\sqrt{n+1}\right)|g,n+1\rangle
\Big].
\end{aligned}
\label{eq:exact_resonant_JC_evolution}
\end{equation}

The origin of the wavepacket splitting is most transparent in the dressed basis
\begin{equation}
|\pm,n\rangle
=
\frac{|e,n\rangle\pm|g,n+1\rangle}{\sqrt{2}},
\qquad
E_{\pm,n}
=
\pm\hbar g\sqrt{n+1}.
\label{eq:JC_dressed_states}
\end{equation}
Since \(
|e,n\rangle=(|+,n\rangle+|-,n\rangle)/\sqrt{2}
\), Eq.~\eqref{eq:exact_resonant_JC_evolution} may be written as
\begin{equation}
|\Psi(t)\rangle
=
\frac{1}{\sqrt{2}}
\sum_{n=0}^{\infty}C_n
\left[
e^{-igt\sqrt{n+1}}|+,n\rangle
+
e^{igt\sqrt{n+1}}|-,n\rangle
\right].
\label{eq:JC_dressed_evolution}
\end{equation}
The two dressed branches thus acquire opposite phases whose rates depend nonlinearly on the magnon number. Dephasing and subsequent rephasing of the Fock components produce the collapse-and-revival dynamics characteristic of the Jaynes--Cummings model.

For a coherent-state distribution concentrated around \(\bar n\gg1\), one may expand
\begin{equation}
\sqrt{n+1}
\simeq
\sqrt{\bar n}
+
\frac{n-\bar n+1}{2\sqrt{\bar n}}
-
\frac{(n-\bar n+1)^2}{8\bar n^{3/2}}
+\cdots .
\label{eq:JC_large_n_expansion}
\end{equation}
The term linear in \(n-\bar n\) generates the leading phase-space rotation, whereas the higher-order terms describe wavepacket deformation and interference. Defining
\begin{equation}
\theta(t)
=
\frac{gt}{2\sqrt{\bar n}},
\label{eq:JC_rotation_angle}
\end{equation}
the factors \(e^{\pm in\theta}\) rotate a coherent state according to
\begin{equation}
e^{\pm i\theta m^\dagger m}|\alpha\rangle
=
|\alpha e^{\pm i\theta}\rangle.
\label{eq:coherent_state_rotation}
\end{equation}
The initial magnon wavepacket therefore bifurcates into two components that counterrotate in phase space \cite{Eiselt1989,Eiselt1991}.

More explicitly, writing \(\alpha=\sqrt{\bar n}\,e^{i\varphi_\alpha}\), the large-\(\bar n\) joint state assumes the approximate form
\begin{equation}
\begin{aligned}
|\Psi(t)\rangle
\simeq
\frac{1}{\sqrt{2}}
\Big[
&
e^{-i\Phi(t)}
|D_+(t)\rangle
\left|\alpha e^{-i\theta(t)}\right\rangle
\\
+&
e^{i\Phi(t)}
|D_-(t)\rangle
\left|\alpha e^{i\theta(t)}\right\rangle
\Big],
\end{aligned}
\label{eq:JC_wavepacket_decomposition}
\end{equation}
where \(\Phi(t)\) is an overall branch phase and
\begin{equation}
|D_\pm(t)\rangle
=
\frac{1}{\sqrt{2}}
\left[
|e\rangle
\pm
e^{-i\varphi_\alpha}
e^{\pm i\theta(t)}
|g\rangle
\right].
\label{eq:JC_atomic_branch_states}
\end{equation}
Equation~\eqref{eq:JC_wavepacket_decomposition} makes explicit that the magnon wavepackets and the associated qubit states evolve together.

The first revival occurs at approximately
\begin{equation}
t_{\rm rev}
\simeq
\frac{2\pi\sqrt{\bar n}}{g}.
\label{eq:JC_revival_time}
\end{equation}
At half of this time,
\begin{equation}
t_{\rm cat}
=
\frac{t_{\rm rev}}{2}
\simeq
\frac{\pi\sqrt{\bar n}}{g},
\qquad
\theta(t_{\rm cat})
\simeq
\frac{\pi}{2},
\label{eq:JC_half_revival_time}
\end{equation}
the two magnon components are centered near \(|-i\alpha\rangle\) and \(|i\alpha\rangle\). Moreover, Eq.~\eqref{eq:JC_atomic_branch_states} gives
\begin{equation}
|D_+(t_{\rm cat})\rangle
=
|D_-(t_{\rm cat})\rangle
\equiv
|D_{\rm A}\rangle,
\label{eq:JC_attractor_state}
\end{equation}
where, up to an overall phase,
\begin{equation}
|D_{\rm A}\rangle
=
\frac{
|e\rangle
+
i e^{-i\varphi_\alpha}|g\rangle
}{\sqrt{2}}.
\end{equation}
The qubit and magnon therefore approximately disentangle, and the joint state factorizes as
\begin{equation}
|\Psi(t_{\rm cat})\rangle
\simeq
|D_{\rm A}\rangle
\otimes
\mathcal{N}
\left[
e^{-i\Phi_{\rm cat}}|-i\alpha\rangle
+
e^{i\Phi_{\rm cat}}|i\alpha\rangle
\right].
\label{eq:JC_half_revival_cat}
\end{equation}
Thus, in the ideal large-\(\bar n\) limit, the resonant dynamics generate a magnonic cat state without requiring qubit postselection \cite{GeaBanacloche1990,GeaBanacloche1991}.

The resonant and dispersive protocols therefore exploit different aspects of the same qubit--magnon interaction, based on already demonstrated experimental findings \cite{Tabuchi2015,LachanceQuirion2017}. In the dispersive regime, the qubit imparts a state-dependent phase rotation to the magnon field. In the resonant regime, the nonlinear \(\sqrt{n+1}\) spectrum of the Jaynes--Cummings ladder causes two dressed wavepackets to counterrotate and recombine into a coherent superposition near half revival.

\section{Parametric Amplification and Gain}
\label{sec:parametric}

Parametric driving makes the cavity sensitive to the phase of the applied drive, thereby modifying both the response and the quantum fluctuations of a cavity–magnon system. In the stable regime, a two-photon pump can amplify and squeeze cavity fluctuations while also changing the response of the hybrid cavity–magnon modes. These effects have been investigated as means of enhancing spin transfer and improving quantum sensing. They also provide a natural route for transferring the parametrically generated cavity fluctuations directly to the magnons.

\subsection{Amplification and active gain}

A two-photon pump coherently creates and annihilates photon pairs, thereby coupling the cavity annihilation and creation operators. In a frame rotating at half the pump frequency, the corresponding interaction can be written as
\begin{equation}
\frac{H_p}{\hbar}
=
\frac{1}{2}
\left(
G_p a^2+G_p^\ast a^{\dagger 2}
\right),
\label{eq:parametric_drive}
\end{equation}
where \(G_p=|G_p|e^{i\phi_p}\) is set by the amplitude and phase of the pump. This interaction underlies both parametrically enhanced spin transfer and squeezing-assisted sensing \cite{MukhopadhyayPRB2022Amplification,WangPRL2026}.

For an isolated parametrically driven cavity, the linear Langevin equations take the form
\begin{equation}
\frac{d}{dt}
\begin{pmatrix}
a\\
a^\dagger
\end{pmatrix}
=
\begin{pmatrix}
-\kappa/2-i\Delta_c & -iG_p^\ast\\
iG_p & -\kappa/2+i\Delta_c
\end{pmatrix}
\begin{pmatrix}
a\\
a^\dagger
\end{pmatrix}
+
\sqrt{\kappa}
\begin{pmatrix}
a_{\rm in}\\
a_{\rm in}^\dagger
\end{pmatrix}.
\label{eq:parametric_cavity_langevin}
\end{equation}
The eigenvalues of this matrix,
\begin{equation}
\mu_\pm
=
-\frac{\kappa}{2}
\pm
\sqrt{|G_p|^2-\Delta_c^2},
\end{equation}
imply the stability condition
\begin{equation}
|G_p|^2
<
\Delta_c^2+\frac{\kappa^2}{4}.
\label{eq:parametric_threshold}
\end{equation}
As the threshold is approached from below, the parametric drive reduces the damping of one cavity quadrature, making it more sensitive to small perturbations. At the same time, fluctuations in the conjugate quadrature are suppressed, leading to squeezing.

\begin{figure}[t]
\centering
\includegraphics[width=0.95\columnwidth]{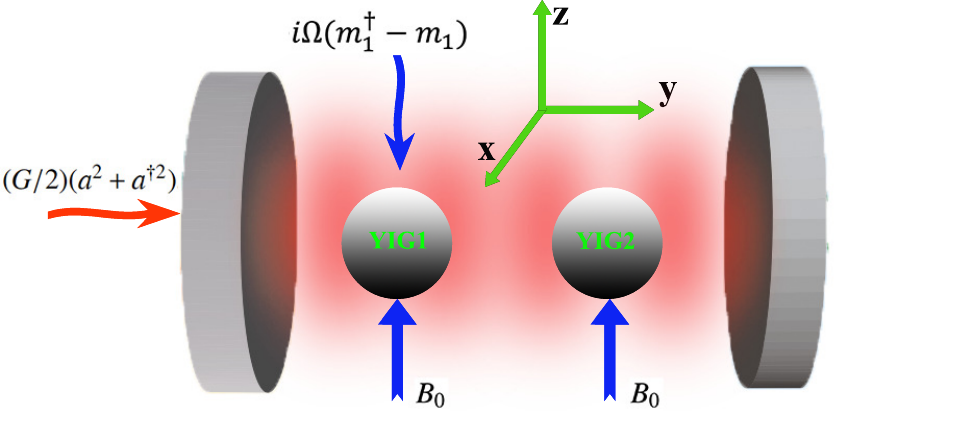}
\caption{Schematic of a parametrically driven cavity coupled to magnon modes. The cavity is driven by a two-photon parametric pump, while the cavity field coherently couples to the magnetic excitations. Taken from Ref.~\cite{MukhopadhyayPRB2022Amplification}.}
    \label{fig:parametric_cavity_magnon}
\end{figure}

More generally, when the cavity is coupled to magnetic modes, this weakly damped quadrature becomes part of a collective magnon--photon excitation. Further, the stability threshold is determined by the full coupled system rather than by the cavity alone. In the basis
\begin{equation}
\mathbf{X}
=
\left(
a,m_1,m_2,
a^\dagger,m_1^\dagger,m_2^\dagger
\right)^T,
\label{eq:nambu_vector}
\end{equation}
the linearized dynamics can be written as
\begin{equation}
\dot{\mathbf{X}}
=
\mathcal{A}\mathbf{X}
+
\Omega\mathbf{F}_{\rm in}
+
\mathbf{N}(t),
\qquad
\mathbf{X}_{\rm ss}
=
-\Omega\mathcal{A}^{-1}\mathbf{F}_{\rm in},
\label{eq:parametric_steady_state}
\end{equation}
where \(\mathcal{A}\) is the Bogoliubov drift matrix, \(\mathbf{F}_{\rm in}\) specifies the coherent drive entering the network, and \(\mathbf{N}(t)\) contains the input-noise operators. A physical steady state exists only when all eigenvalues of \(\mathcal{A}\) have negative real parts.

The frequency-dependent susceptibility can be defined as the Green's function $\boldsymbol{\chi}(\omega)
=
\left(
-i\omega\mathbb{I}-\mathcal{A}
\right)^{-1}$. Parametric enhancement arises when the damping rate of one collective Bogoliubov mode approaches zero from within the stable region, producing a large susceptibility. This mechanism directly enhances cavity-mediated spin transport. Consider two ferrimagnetic samples coupled to a common cavity, with the first magnon mode driven coherently. The cavity transfers part of this excitation to the second sample, where the resulting magnon population generates an electrically detectable spin-pumping signal. For a large coherent drive, the spin current goes as
\begin{equation}
I_s^{(2)}
\propto
\langle m_2^\dagger m_2\rangle
\simeq
|\bar m_2|^2,
\qquad
\bar m_2
=
-\Omega
\left[
\mathcal{A}^{-1}\mathbf{F}_{\rm in}
\right]_{m_2}.
\label{eq:parametric_spin_current}
\end{equation}
Because the two-photon drive modifies the susceptibility matrix \(\mathcal{A}^{-1}\), it can enhance the remote spin current by several orders of magnitude close to the stability boundary \cite{MukhopadhyayPRB2022Amplification}.

This amplification can be understood in two ways. In the Fock-state picture, the two-photon pump couples states whose cavity occupations differ by two, thereby opening higher-order paths through the cavity--magnon excitation ladder. Alternatively, in the collective-mode picture, the same interaction produces a long-lived hybrid mode whose reduced damping strongly amplifies the steady-state response \cite{MukhopadhyayPRB2022Amplification}.

We anticipate that the sensing potential for a long-lived mode discussed earlier in Sec. \ref{sec:dissipative} can be further enhanced through parametric driving. While dissipatively coupled anti-$\mathcal{PT}$ systems ~\cite{Wang2022QFI} yield a sensitivity of $\delta\varepsilon_{\rm CRB}\propto\varepsilon^2$, recent work on parametrically driven systems with gain points to stronger precision scaling. At an $n$th-order EP, it has been shown that
\begin{equation}
\mathcal{F}_{Q}(\theta)\propto\theta^{-4n},
\qquad
\delta\theta_{\rm CRB}\propto\theta^{2n},
\label{eq:ep_squeezing_scaling}
\end{equation}
where $\theta$ is the parameter to be estimated \cite{WangPRL2026}. Thus, a second-order EP would imply $\delta\theta_{\rm CRB}\propto\theta^4$. This enhancement reflects the combined response of the non-Hermitian mode structure and the squeezed-state fluctuations. Moreover, it suggests an analogous sensing enhancement via long-lived modes in dissipatively coupled systems.

Parametric driving can also reorganize the nonequilibrium phase structure of cavity-magnonic systems. It has been shown that driven cavity--magnon platforms may undergo first-or second-order parity-symmetry-breaking transitions between phases with microscopic and macroscopic occupations \cite{Zhang2021Parametric}, linking parametric mode mixing to nonlinear stability and critical dynamics.

An alternative to parametric driving is the direct injection of gain into the cavity. This regime has been realized experimentally using a microwave resonator containing an active amplification element, so that amplified photons hybridize with magnons to form gain-driven polaritons \cite{Yao2023Gain}. The resulting dynamics include self-sustained polariton oscillations, bright-mode selection, magnon--photon synchronization, coherent microwave amplification, and narrow-linewidth emission. Parametric driving therefore provides phase-sensitive gain through mode mixing, whereas an active cavity sustains collective oscillations through net gain.

Parametric driving and direct gain thus provide complementary routes to controlling cavity--magnon systems far from equilibrium. The former reshapes the hybrid susceptibility and quantum fluctuations, while the latter can sustain coherent polariton dynamics. In both cases, useful enhancement is ultimately constrained by stability, bandwidth, saturation, response time, and noise.

\subsection{Future perspective: Parametric transfer of squeezing to magnons}
\label{sec:parametric-b}

The two-photon parametric interaction used to amplify spin transport~\cite{MukhopadhyayPRB2022Amplification} can be utilized for the generation of squeezed microwave fluctuations that are transferred directly to the collective magnon modes. This provides an alternative to earlier squeezing-transfer schemes that rely on externally generated squeezed microwave fields \cite{Li2019Squeezing,Nair2020Deterministic}.

Consider two identical magnon modes with $g_1=g_2=g_0$ and $\Delta_1=\Delta_2=\Delta_m$. Defining the bright and dark modes as $M=(m_1+m_2)/\sqrt{2}$ and $D=(m_1-m_2)/\sqrt{2}$, respectively, only $M$ couples to the cavity, with collective coupling $g_B=\sqrt{2}g_0$, while $D$ remains dark. Choosing the pump phase such that $G_p>0$, the relevant fluctuation Hamiltonian is
\begin{equation}
\frac{H_{\rm fl}}{\hbar}
=
\Delta_c a^\dagger a
+
\Delta_m M^\dagger M
+
g_B(a^\dagger M+aM^\dagger)
+
\frac{iG_p}{2}
\left(a^{\dagger 2}-a^2\right).
\label{eq:future_magnon_squeezing}
\end{equation}

Thus, the subsequent discussion readily applies to a single magnon as well. The squeezing transfer is especially easy to see on resonance, $\Delta_c=\Delta_m=0$, where the quadrature dynamics separate into two independent sectors. With the above pump-phase convention, the parametric interaction squeezes the cavity quadrature $P_a=(a-a^\dagger)/(i\sqrt{2})$ while amplifying its conjugate $X_a$. The cavity--magnon beam-splitter interaction couples $P_a$ directly to the bright-magnon quadrature $X_M=(M+M^\dagger)/\sqrt{2}$. The $(P_a,X_M)$ sector therefore directly describes the transfer of cavity squeezing to the magnons. The Langevin equations for these quadratures are given by
\begin{align}
\dot P_a
&=
-\left(\frac{\kappa}{2}+G_p\right)P_a
-g_BX_M
+\sqrt{\kappa}\,P_a^{\rm in},
\\
\dot X_M
&=
-\frac{\gamma}{2}X_M
+g_BP_a
+\sqrt{\gamma}\,X_M^{\rm in},
\label{eq:future_quadratures}
\end{align}
where $\kappa$ and $\gamma$ are the cavity and magnon linewidths, respectively.

To include thermal magnon fluctuations, we take the microwave input channels to be effectively in vacuum while characterizing the magnon bath by a thermal occupation $\bar n_m=[\exp(\hbar\omega_m/k_BT)-1]^{-1}$. The input noises are assumed to be Markovian, with zero mean and symmetrized correlations
\begin{align}
\frac{1}{2}
\left\langle
\left\{
P_a^{\rm in}(t),P_a^{\rm in}(t')
\right\}
\right\rangle
&=
\frac{1}{2}\delta(t-t'),
\\
\frac{1}{2}
\left\langle
\left\{
X_M^{\rm in}(t),X_M^{\rm in}(t')
\right\}
\right\rangle
&=
\left(\bar n_m+\frac{1}{2}\right)\delta(t-t'),
\end{align}
while the cavity and magnon input noises are mutually uncorrelated, $\left\langle
P_a^{\rm in}(t)X_M^{\rm in}(t')
\right\rangle
=
0$. For two identical and independent magnon reservoirs with the same thermal occupation, the bright-mode noise $X_M^{\rm in}$ inherits the same thermal variance.

We define, for brevity, $K_+=\kappa/2+G_p$, $V_a=\langle P_a^2\rangle$, $V_M=\langle X_M^2\rangle$, and $C=\frac{1}{2}\langle P_aX_M+X_MP_a\rangle$. Using the above noise correlations, the Langevin equations yield
\begin{align}
\dot V_a
&=
-2K_+V_a-2g_BC+\frac{\kappa}{2},
\\
\dot V_M
&=
-\gamma V_M+2g_BC
+\gamma\left(\bar n_m+\frac{1}{2}\right),
\\
\dot C
&=
-\left(K_++\frac{\gamma}{2}\right)C
+g_B(V_a-V_M).
\label{eq:future_second_moments}
\end{align}

Solving these equations in the steady state gives
\begin{align}
\left\langle\Delta X_M^2\right\rangle_T
&=
\frac{1}{2}
-
\frac{
g_B^2G_p
}{
2\left(K_++\gamma/2\right)
\left(K_+\gamma/2+g_B^2\right)
}
\notag\\
&\quad+
\bar n_m
\frac{
K_+ + g_B^2/(K_++\gamma/2)
}{
K_+ + 2g_B^2/\gamma
}.
\label{eq:future_magnon_variance_T}
\end{align}
The second term describes squeezing transferred from the cavity, whereas the final term represents the competing thermal contribution. Thermal fluctuations are generally detrimental to squeezing because they increase the quadrature variance above the vacuum level. Here, however, their effect is partially suppressed by the cavity coupling. Since $g_B^2/(K_++\gamma/2)<2g_B^2/\gamma$, the factor multiplying $\bar n_m$ is smaller than unity. The cavity, therefore, not only transfers squeezing to the magnon mode but also reduces its sensitivity to thermal noise, allowing the squeezing to remain observable at finite temperature. Specifically, sub-vacuum magnon squeezing requires
\begin{equation}
\bar n_m
<
\bar n_m^{\rm crit}
\equiv
\frac{g_B^2G_p}
{\gamma\left[
K_+\left(K_++\gamma/2\right)+g_B^2
\right]}.
\label{eq:future_thermal_squeezing_condition}
\end{equation}
Thus, squeezing survives as long as the thermal occupation remains below the critical value $\bar n_m^{\rm crit}$. At zero temperature ($\bar n_m=0$), Eq.~\eqref{eq:future_magnon_variance_T} reduces to
\begin{equation}
\left\langle\Delta X_M^2\right\rangle_{T=0}
=
\frac{1}{2}
-
\frac{
g_B^2G_p
}{
2\left(K_++\gamma/2\right)
\left(K_+\gamma/2+g_B^2\right)
}.
\label{eq:future_magnon_variance_zeroT}
\end{equation}
To make the dependence on the parametric drive explicit, we define $x=2G_p/\kappa$, $\alpha=1+\gamma/\kappa$, and $\beta=1+\mathcal{C}_B$, where $\mathcal{C}_B=4g_B^2/(\kappa\gamma)$ is the cooperativity. The magnitude of the squeezing contribution can then be rewritten as
\begin{equation}
S_0
=
\frac{\mathcal{C}_B}{2}
\frac{x}{(x+\alpha)(x+\beta)}.
\end{equation}
As a function of $x$, $S_0$ reaches its algebraic maximum at $x=\sqrt{\alpha\beta}$. However, dynamical stability requires $x<\min(\alpha,\beta)$. Since $\min(\alpha,\beta)\leq\sqrt{\alpha\beta}$, the entire stable regime lies on the increasing branch of $S_0$. Thus, the transferred magnon squeezing increases monotonically with $G_p$ throughout the stable regime. The collective cooperativity $\mathcal{C}_B$ controls the efficiency of this transfer, with larger cooperativity producing stronger magnon squeezing for a given parametric drive.

We also note that since $X_M=(X_1+X_2)/\sqrt{2}$, sub-vacuum bright-mode fluctuations correspond to squeezing of a joint magnon quadrature. At zero temperature, the decoupled dark mode remains in vacuum, so the cavity-generated squeezing produces nonclassical two-magnon correlations. This mechanism complements direct magnetic parametric pumping, where single- and two-mode thermal magnon squeezing has recently been observed \cite{Hioki2026}.

\section{Outlook}
\label{sec:outlook}

Cavity magnonics has steadily expanded beyond the study of coherent magnon--photon polaritons to include engineered dissipation, nonlinear and parametric effects, and quantum-state preparation. Recent progress in generating squeezed magnon states, together with proposals for magnon--magnon entanglement, has opened new possibilities for preparing and controlling more complex nonclassical states of collective spin excitations.

Superconducting qubits offer a promising route toward this goal. Both coherent and strongly dispersive qubit--magnon coupling have been demonstrated \cite{Tabuchi2015,LachanceQuirion2017}, providing the key ingredients for creating magnonic Schr"odinger-cat states, as identified earlier in this review. Realizing such states would extend quantum magnonics beyond Gaussian-state preparation and enable greater quantum control over collective magnetic excitations.

Cavity-mediated interactions also provide a way to control spin transport between spatially separated magnetic systems, which has been experimentally demonstrated \cite{Bai2017}. Intrinsic magnetic nonlinearities could add another level of control to such processes. In addition, intrinsic magnetic nonlinearities to magnetocrystalline anisotropy give rise to the magnon Kerr effect \cite{Wang2016Kerr}, which can lead to bistability of cavity--magnon polaritons under strong driving \cite{Wang2018Bistability}. In multimode systems, Kerr nonlinearities and cavity-mediated interactions can produce multiple coexisting steady states with different spin-current states, also termed as multistability \cite{Nair2020SpinCurrents,Shen2021Multistability}. Controlled switching between such states could provide new possibilities for multilevel magnetic signal processing. An important next step would be to understand how quantum fluctuations affect switching, metastability, and the selection between these states, bringing this nonlinear behavior into the quantum regime.

Dissipative coupling presents a unique resource for modifying the collective response. Correlated losses can redistribute damping among the hybrid modes, creating, in particular, long-lived excitations with enhanced susceptibility \cite{Harder2018,Nair2021Sensing,Nair2021Ultralow}. These modes must, however, be accessible for excitation and readout, so care must be taken to balance longer lifetimes with sufficient coupling to the external channels. Reservoir engineering can also provide new ways to achieve directional transport. Studies of waveguide-based systems have shown that nonreciprocity can be enhanced through bound states in the continuum, nonlinear switching, and chiral light--matter interactions \cite{Biehs2023,Miao2024,Miao2025}. Applying these ideas to cavity-magnonic platforms could offer new ways to control the direction of microwave and spin transport through engineered dissipation, magnetic nonlinearities, and quantum fluctuations. Finally, Floquet modulation can add a new dimension to magnonic control by enabling phase-selective coupling and synthetic interference, while combining periodic driving with engineered dissipation to achieve tunable nonreciprocal magnonic transport. 

Parametric driving and gain provide complementary techniques to enhance the response of cavity--magnon systems. Near the instability threshold, parametric driving can amplify spin transfer, modify the hybrid-mode response, and generate squeezed states \cite{MukhopadhyayPRB2022Amplification,WangPRL2026}. An especially promising direction is the transfer of parametrically induced cavity squeezing to collective magnon modes, providing an in-situ route to nonclassical magnon fluctuations without requiring intrinsic magnetic nonlinearity, as outlined in Sec.\ref{sec:parametric-b}. In practice, these enhancements need to be balanced against stability, bandwidth, saturation, and added noise.

Precision sensing provides a natural application for these enhancement mechanisms. EPs, reservoir-induced coherence, parametric squeezing, and dissipative criticality can all enhance the response to weak perturbations \cite{Nair2021Sensing,WangPRL2026}. A particularly attractive possibility is to combine the enhanced response of dissipatively engineered long-lived modes with parametric squeezing, which could further improve the precision scaling discussed in Sec.~\ref{sec:dissipative}. An important future direction is to determine whether this enhanced response translates into improved precision once noise, state preparation limitations, and readout are properly taken into account. Addressing these challenges could open new directions for cavity magnonics in quantum sensing and information processing.

\begin{acknowledgments}
The authors thank C. M. Hu for many fruitful discussions over the years. G.S.A. acknowledges support from the Welch Foundation under Grant No.~1943 and the National Science Foundation under Award No.~2426699.
\end{acknowledgments}

\section*{Author Declarations}

\subsection*{Conflict of Interest}

The authors have no conflicts to disclose.

\subsection*{Author Contributions}

J.M.P.N. \& D.M.: Conceptualization and writing. 

G.S.A.: Conceptualization, supervision, review, \& editing.

J.M.P.N and D.M. contributed equally to the manuscript.

\section*{Data Availability}

Data sharing does not apply to this article as no new data were created or analyzed in this study.

\bibliography{references}

\end{document}